\documentclass[10pt]{article}
\usepackage{amsthm,amsmath,amssymb}
\usepackage{subfig,sidecap,caption}
\usepackage{paralist}
\usepackage[usenames,dvipsnames]{color}
\usepackage[pdftex,breaklinks,colorlinks,
    citecolor={BlueViolet}, linkcolor={Blue},urlcolor=Maroon]{hyperref}  
\usepackage{tikz,pgfkeys}
\usepackage{tkz-graph}
\usetikzlibrary{decorations.pathmorphing}
\usetikzlibrary{quotes}
\usetikzlibrary{arrows.meta}
\usetikzlibrary{arrows,shapes}
\usetikzlibrary{decorations.pathreplacing, decorations.shapes}
\usepackage{graphicx}
\usepackage{charter,eulervm}%
\usepackage{multirow,booktabs,array}
\usepackage{tabularx}
\usepackage[final,expansion=alltext,protrusion=true]{microtype}
\usepackage{float}

\theoremstyle{plain} 
\newtheorem{theorem}{Theorem}[section]

\newtheorem{proposition}[theorem]{Proposition}
\theoremstyle{definition}
\newtheorem{definition}{Definition}[section]

\newcommand{\hsc}[1]{{\footnotesize\sf\MakeUppercase{#1}}}
\newcommand{\problem}[2]{\hsc{#1} $\rightarrow$ \hsc{#2}}
\newcommand{\lp}[1]{\ensuremath{{\mathtt{lp}(#1)}}}
\newcommand{\rp}[1]{\ensuremath{{\mathtt{rp}(#1)}}}
\newcommand{\comment}[1]{\textbackslash\!\!\textbackslash {\em #1}}
\tikzstyle{class} = [shape=rectangle, rounded corners, draw, align=center, top color=white, bottom color=blue!20]
\tikzstyle{vertex}  = [{fill=blue,circle,draw,inner sep=1pt}]

\title{Solving 3SAT By Reduction To Testing For Odd Hole}

\author{M. Delacorte}

\date{}

\begin{document}
\maketitle

\begin{abstract}
  An algorithm is given for finding the solutions to 3SAT problems.  The algorithm uses Bienstock's reduction from 3SAT to existence of induced odd cycle of length greater than three, passing through a prescribed node in the constructed graph.  The algorithm  proceeds to find what will be called the hole complexes of the graph.  The set of the boundary nodes of the hole complex containing the prescribed node is then searched for the subsets of 8 nodes corresponding to the 3SAT's literals.  If a complete set of literals is contained in the boundary then the 3SAT is solvable.
  
\end{abstract}

\section{Introduction}\label{sec:intro}
Bienstock [1] gives a reduction from 3SAT to  existence of induced odd cycle of length greater than three, passing through a prescribed node $u$ in the constructed graph [1].  For a given 3SAT we have $n$ variables $x_1, . . . , x_n$, and $m$ clauses $c_1, . . . , c_m$.  Each clause is denoted as $(z_1 \lor z_2 \lor z_3)$ where $z_i = x_j or \overline x_j$.  The construction uses two gadgets (see Fig. 1) one for clauses the other for variables with color coded edges $blue$ and $red$ (red in bold).  The red edges are the constraints in the problem.  Three additional nodes are used in the construction $u$ the prescribed node to be in the odd hole and $w$ and $v$. The graph is completed with the following edges:
        	\begin{flushleft}
            These edges in blue.
            ${u, w}$, ${u, c_{1,2}}$ ${w, c_{t,1}}$; for $1 \leq i < n{c_{i,3}, c_{i+l,1}}$ and ${c_{i,4}, c_{i+l,2}}$; 
			${c_{n,3}, d_{1,1}}$ ${c_{n,4}, d_{1,2}}$; for$1 \leq j < n{d_{j,3}, d_{j+l,1}}$  and ${d_{j,4}, d_{j+l,2}}$; and ${d_{m,3}, v}, {d_{m,4}, v}$.
			 \end{flushleft}
			\begin{flushleft}
			 In addition, $G$ contains, for each clause $C_j$, the following red edges. Let $C_j = (z_1 \lor z_2, \lor z_3)$. For $1 \leq k \leq 3$ if $z_k = x_i$, (some $i$) then we have the edges ${f_{i,1}, f_j(z_k)}$ and ${f_{i,3}, f_j(z_k)}$ while if $z_k = \overline x_i$ the edges are ${t_{i,1}, f_j(z_k)}$ and ${t_{i,3}, f_j(z_k)}$.
			 \end{flushleft}
\begin{flushleft}
(see fig. 2, 3, 4) showing the graphs for the following formula.		 
$(x_1+x_2+x_3)(x_1+x_2+\overline x_3)(x_1+\overline x_2+x_3)(x_1+\overline x_2+\overline x_3)(\overline x_1+x_2+x_3)(\overline x_1+x_2+\overline x_3)(\overline x_1+\overline x_2+x_3)(\overline x_1+\overline x_2+\overline x_3)$		 
Figure 2 shows the gadget version, figure 3 the untwisted graph and figure 4 the graph without constraints.	
\end{flushleft}	 
Bienstock then proves the following.
			 
	\begin{proposition}			
		Let L be an odd hole containing u.  Then, for $1 \leq i \leq n$, exactly one of the following is true:
	  \begin{enumerate}[(1)]
	  \item L contains the blue paths $c_{i,1}$, $t_{i,1}$, $c_{i,3}$ and $c_{i,2}$, $t_{i,2}$, $t_{i,3}$, $t_{i,4}$, $c_{i,4}$ 
	  \item L contains the paths $c_{i,1}$, $f_{i,1}$, $c_{i,3}$ and $c_{i,2}$, $f_{i,2}$, $f_{i,3}$, $f_{i,4}$, $c_{i,4}$.
      \end{enumerate}
    \end{proposition}

\section{Hole Complexes}\label{sec:holes}

\begin{definition}{Hole Complex}
	A hole complex is a set of path graphs length $\geq2$ with their end nodes joined to nodes in other paths.  The result must be a connected graph with each path part of at least one hole and any two holes must be linked by a chain of holes where each pair of linked holes in the chain share at least two non adjacent nodes.(see Fig. 5)
\end{definition}

   In a graph we can consider the individual holes (cordless cycles size $>3$) or the hole complexes.  A single hole may be a complex. Two holes
   with two or more nodes in common (if only two nodes they must be non adjacent) form a hole complex. In a graph a hole complex to which no further holes in the graph can be added will be called a maximal hole complex of the graph. A hole complex can have an exponential number of holes in the number of its nodes [2]. One can find the maximal hole complexes of a graph by using a hole detecting algorithm [3]. After finding a hole its nodes are recorded and then it is filled by adding its anti hole edges to the graph (see Fig. 6). This action is repeated till no more holes are detected. Note some of the holes will contain added anti hole edges (see Fig. 7) these will be removed from the hole leaving path graphs.  This procedure will give the path graphs of the hole complexes of the graph with the possible absence of some paths of length two (see Fig. 6). To insure the inclusion of all length two paths search the original graph for paths of length two not part of triangles and with end node degrees greater than two. To find all the maximal hole complexes pick a hole or set of paths from the list of all the holes and paths found above along with the two paths search the remaining holes and paths for any that share at least two non adjacent nodes with the hole or path(s) add any that do to the maximal hole list and remove them from the hole list. Now search the list of holes and paths for holes and paths that share nodes with the newly added holes or paths. Continue this process till no more holes and paths are found to add to the maximal hole list. Repeat the above till all the maximal hole complexes have been found and the list of holes and paths is exhausted.

\section{Finding solutions}\label{sec:3SAT}

Given a 3SAT problem of $n$ variables apply Bienstock's reduction to the problem to obtain a graph with holes.  Find the maximal hole complexes of the graph as outlined above.  Among the maximal hole complexes find the one containing the prescribed node $u$.  Search its nodes for the sets $c_{i,1}$, $t_{i,1}$, $c_{i,3}$ $c_{i,2}$, $t_{i,2}$, $t_{i,3}$, $t_{i,4}$, $c_{i,4}$ and  $c_{i,1}$, $f_{i,1}$, $c_{i,3}$ $c_{i,2}$, $f_{i,2}$, $f_{i,3}$, $f_{i,4}$, $c_{i,4}$ if one or both of these sets is present for each i of the n variables then the 3SAT is solvable.

\begin{figure}[H]
	\centering\small
	\begin{tikzpicture}[scale = 1.0]
	
	
	\tikzstyle{every node}=[circle,draw=black,fill=white,thick,minimum size=6pt,
	inner sep=0pt]

	\draw (3.5,0.5) node (fi-2) [label=below:$f_{i,2}$] {};
	\draw (4.5,0.5) node (fi-3) [label=below:$f_{i,3}$] {};
	\draw (5.5,0.5) node (fi-4) [label=below:$f_{i,4}$] {};
	\draw (2,1.5) node (ci-2) [label=below:$c_{i,2}$] {};
	\draw (7,1.5) node (ci-4) [label=below:$c_{i,4}$] {};
	\draw (3.5,2.5) node (ti-2) [label=below:$t_{i,2}$] {};
	\draw (4.5,2.5) node (ti-3) [label=below:$t_{i,3}$] {};
	\draw (5.5,2.5) node (ti-4) [label=below:$t_{i,4}$] {};
	\draw (3.5,3.5) node (fi-1) [label=left:$f_{i,1}$] {};
	\draw (2,4.5) node (ci-1) [label=below:$c_{i,1}$] {};
	\draw (5,4.5) node (ci-3) [label=below:$c_{i,3}$] {};
	\draw (3.5,5.5) node (ti-1) [label=below:$t_{i,1}$] {};

	\draw (13.5,1) node (fjz3) [label=below:$f_j({z}_3)$] {};
	\draw (12,1.5) node (dj-2) [label=below:$d_{j,2}$] {};
	\draw (13.5,1.5) node (fjz2) [label=right:$f_j({z}_2)$] {};
	\draw (15,1.5) node (dj-4) [label=below:$d_{j,4}$] {}; 
	\draw (13.5,2) node (fjz1) [label=left:$f_j({z}_1)$] {};
	\draw (12,2.5) node (dj-1) [label=below:$d_{j,1}$] {};
	\draw (13.5,2.5) node (rj) [label=right:$r_j$] {};
	\draw (15,2.5) node (dj-3) [label=below:$d_{j,3}$] {};

	\begin{scope}[red,very thick,rounded corners]
	\draw (fi-2)--(1,0.5)--(1,5.5)--(ti-1)--(8,5.5)--(8,0)--(5,0)--(fi-3)--(ti-3)--(fi-1)--(ti-2);
	\end{scope}
	
	\begin{scope}[blue]
	\draw (ti-1)--(ci-3)--(fi-1)--(ci-1)--(ti-1);
    \draw (ti-2)--(ti-3)--(ti-4)--(ci-4)--(fi-4)--(fi-3)--(fi-2)--(ci-2)--(ti-2);
    \draw (fi-2)--(fi-3);
	\end{scope}
	
	\begin{scope}[blue]
    \draw (dj-1)--(rj)--(dj-3);
    \draw (dj-4)--(fjz3)--(dj-2)--(fjz2)--(dj-4)--(fjz1)--(dj-2);
	\end{scope}
	
	\end{tikzpicture}
	\caption{Gadgets used in reduction from $3SAT$ to odd hole containing specified node.  a) variable gadget b) clause gadget }
	\label{fig:overview}
\end{figure}
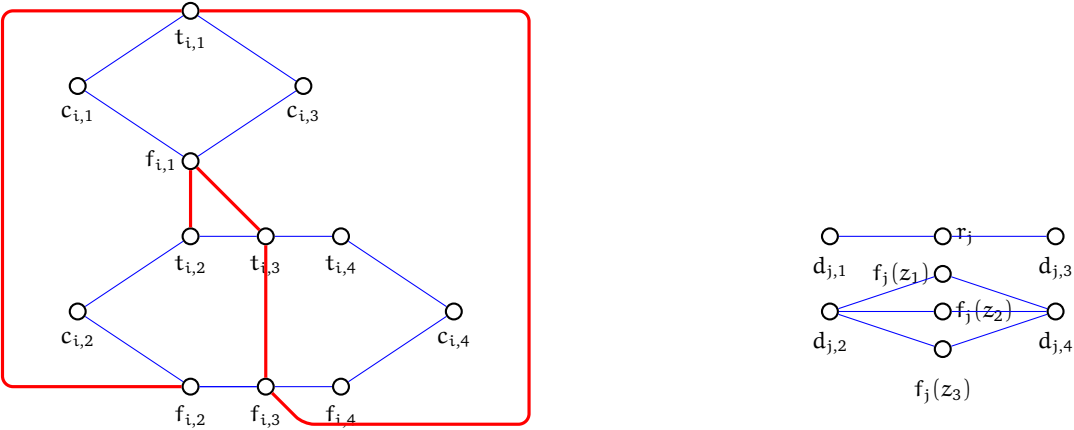

\begin{figure}[H]
	\centering\small
\begin{tikzpicture}


\tikzstyle{every node}=[circle,draw=black,fill=white,thick,minimum size=6pt,
inner sep=0pt]

\draw (0,16.5) node (1) [label=right:$w$] {};
\draw (0,13.5) node (2) [label=right:$u$] {};
\draw (16,2) node (3) [label=right:$v$] {};

\draw (3.5,0.5) node (f3-2) [label=below:$f_{3,2}$] {};
\draw (4.5,0.5) node (f3-3) [label=below:$f_{3,3}$] {};
\draw (5.5,0.5) node (f3-4) [label=below:$f_{3,4}$] {};
\draw (2,1.5) node (c3-2) [label=below:$c_{3,2}$] {};
\draw (7,1.5) node (c3-4) [label=below:$c_{3,4}$] {};
\draw (3.5,2.5) node (t3-2) [label=below:$t_{3,2}$] {};
\draw (4.5,2.5) node (t3-3) [label=below:$t_{3,3}$] {};
\draw (5.5,2.5) node (t3-4) [label=below:$t_{3,4}$] {};
\draw (3.5,3.5) node (f3-1) [label=left:$f_{3,1}$] {};
\draw (2,4.5) node (c3-1) [label=below:$c_{3,1}$] {};
\draw (5,4.5) node (c3-3) [label=below:$c_{3,3}$] {};
\draw (3.5,5.5) node (t3-1) [label=below:$t_{3,1}$] {};

\draw (3.5,6.5) node (f2-2) [label=left:$f_{2,2}$] {};
\draw (4.5,6.5) node (f2-3) [label=below:$f_{2,3}$] {};
\draw (5.5,6.5) node (f2-4) [label=below:$f_{2,4}$] {};
\draw (2,7.5) node (c2-2) [label=below:$c_{2,2}$] {};
\draw (7,7.5) node (c2-4) [label=below:$c_{2,4}$] {};
\draw (3.5,8.5) node (t2-2) [label=below:$t_{2,2}$] {};
\draw (4.5,8.5) node (t2-3) [label=below:$t_{2,3}$] {};
\draw (5.5,8.5) node (t2-4) [label=below:$t_{2,4}$] {};
\draw (3.5,9.5) node (f2-1) [label=left:$f_{2,1}$] {};
\draw (2,10.5) node (c2-1) [label=below:$c_{2,1}$] {};
\draw (5,10.5) node (c2-3) [label=below:$c_{2,3}$] {};
\draw (3.5,11.5) node (t2-1) [label=below:$t_{2,1}$] {};

\draw (3.5,12.5) node (f1-2) [label=left:$f_{1,2}$] {};
\draw (4.5,12.5) node (f1-3) [label=below:$f_{1,3}$] {};
\draw (5.5,12.5) node (f1-4) [label=below:$f_{1,4}$] {};
\draw (2,13.5) node (c1-2) [label=below:$c_{1,2}$] {};
\draw (7,13.5) node (c1-4) [label=below:$c_{1,4}$] {};
\draw (3.5,14.5) node (t1-2) [label=below:$t_{1,2}$] {};
\draw (4.5,14.5) node (t1-3) [label=below:$t_{1,3}$] {};
\draw (5.5,14.5) node (t1-4) [label=below:$t_{1,4}$] {};
\draw (3.5,15.5) node (f1-1) [label=left:$f_{1,1}$] {};
\draw (2,16.5) node (c1-1) [label=below:$c_{1,1}$] {};
\draw (5,16.5) node (c1-3) [label=below:$c_{1,3}$] {};
\draw (3.5,17.5) node (t1-1) [label=below:$t_{1,1}$] {};

\draw (13.5,1) node (f8z3) [label=below:$f_8({z}_3)$] {};
\draw (12,1.5) node (d8-2) [label=below:$d_{8,2}$] {};
\draw (13.5,1.5) node (f8z2) [label=right:$f_8({z}_2)$] {};
\draw (15,1.5) node (d8-4) [label=below:$d_{8,4}$] {}; 
\draw (13.5,2) node (f8z1) [label=left:$f_8({z}_1)$] {};
\draw (12,2.5) node (d8-1) [label=below:$d_{8,1}$] {};
\draw (13.5,2.5) node (r8) [label=right:$r_8$] {};
\draw (15,2.5) node (d8-3) [label=below:$d_{8,3}$] {};

\draw (13.5,3) node (f7z3) [label=right:$f_7({z}_3)$] {};
\draw (12,3.5) node (d7-2) [label=below:$d_{7,2}$] {};
\draw (13.5,3.5) node (f7z2) [label=right:$f_7({z}_2)$] {};
\draw (15,3.5) node (d7-4) [label=below:$d_{7,4}$] {}; 
\draw (13.5,4) node (f7z1) [label=left:$f_7({z}_1)$] {};
\draw (12,4.5) node (d7-1) [label=below:$d_{7,1}$] {};
\draw (13.5,4.5) node (r7) [label=right:$r_7$] {};
\draw (15,4.5) node (d7-3) [label=below:$d_{7,3}$] {}; 

\draw (13.5,5) node (f6z3) [label=right:$f_6({z}_3)$] {};
\draw (12,5.5) node (d6-2) [label=below:$d_{6,2}$] {};
\draw (13.5,5.5) node (f6z2) [label=right:$f_6({z}_2)$] {};
\draw (15,5.5) node (d6-4) [label=below:$d_{6,4}$] {}; 
\draw (13.5,6) node (f6z1) [label=left:$f_6({z}_1)$] {};
\draw (12,6.5) node (d6-1) [label=below:$d_{6,1}$] {};
\draw (13.5,6.5) node (r6) [label=right:$r_6$] {};
\draw (15,6.5) node (d6-3) [label=below:$d_{6,3}$] {}; 

\draw (13.5,7) node (f5z3) [label=right:$f_5({z}_3)$] {};
\draw (12,7.5) node (d5-2) [label=below:$d_{5,2}$] {};
\draw (13.5,7.5) node (f5z2) [label=right:$f_5({z}_2)$] {};
\draw (15,7.5) node (d5-4) [label=below:$d_{5,4}$] {}; 
\draw (13.5,8) node (f5z1) [label=left:$f_5({z}_1)$] {};
\draw (12,8.5) node (d5-1) [label=below:$d_{5,1}$] {};
\draw (13.5,8.5) node (r5) [label=right:$r_5$] {};
\draw (15,8.5) node (d5-3) [label=below:$d_{5,3}$] {}; 

\draw (13.5,9) node (f4z3) [label=right:$f_4({z}_3)$] {};
\draw (12,9.5) node (d4-2) [label=below:$d_{4,2}$] {};
\draw (13.5,9.5) node (f4z2) [label=right:$f_4({z}_2)$] {};
\draw (15,9.5) node (d4-4) [label=below:$d_{4,4}$] {}; 
\draw (13.5,10) node (f4z1) [label=left:$f_4({z}_1)$] {};
\draw (12,10.5) node (d4-1) [label=below:$d_{4,1}$] {};
\draw (13.5,10.5) node (r4) [label=right:$r_4$] {};
\draw (15,10.5) node (d4-3) [label=below:$d_{4,3}$] {}; 

\draw (13.5,11) node (f3z3) [label=right:$f_3({z}_3)$] {};
\draw (12,11.5) node (d3-2) [label=below:$d_{3,2}$] {};
\draw (13.5,11.5) node (f3z2) [label=right:$f_3({z}_2)$] {};
\draw (15,11.5) node (d3-4) [label=below:$d_{3,4}$] {}; 
\draw (13.5,12) node (f3z1) [label=left:$f_3({z}_1)$] {};
\draw (12,12.5) node (d3-1) [label=below:$d_{3,1}$] {};
\draw (13.5,12.5) node (r3) [label=right:$r_3$] {};
\draw (15,12.5) node (d3-3) [label=below:$d_{3,3}$] {}; 

\draw (13.5,13) node (f2z3) [label=right:$f_2({z}_3)$] {};
\draw (12,13.5) node (d2-2) [label=below:$d_{2,2}$] {};
\draw (13.5,13.5) node (f2z2) [label=right:$f_2({z}_2)$] {};
\draw (15,13.5) node (d2-4) [label=below:$d_{2,4}$] {}; 
\draw (13.5,14) node (f2z1) [label=left:$f_2({z}_1)$] {};
\draw (12,14.5) node (d2-1) [label=below:$d_{2,1}$] {};
\draw (13.5,14.5) node (r2) [label=right:$r_2$] {};
\draw (15,14.5) node (d2-3) [label=below:$d_{2,3}$] {}; 

\draw (13.5,15) node (f1z3) [label=right:$f_1({z}_3)$] {};
\draw (12,15.5) node (d1-2) [label=below:$d_{1,2}$] {};
\draw (13.5,15.5) node (f1z2) [label=right:$f_1({z}_2)$] {};
\draw (15,15.5) node (d1-4) [label=below:$d_{1,4}$] {}; 
\draw (13.5,16) node (f1z1) [label=left:$f_1({z}_1)$] {};
\draw (12,16.5) node (d1-1) [label=below:$d_{1,1}$] {};
\draw (13.5,16.5) node (r1) [label=right:$r_1$] {};
\draw (15,16.5) node (d1-3) [label=below:$d_{1,3}$] {};

\begin{scope}[red,very thick,rounded corners]
\draw (f3-2)--(1,0.5)--(1,5.5)--(t3-1)--(8,5.5)--(8,0)--(5,0)--(f3-3)--(t3-3)--(f3-1)--(t3-2);
\draw (f2-2)--(1,6.5)--(1,11.5)--(t2-1)--(8,11.5)--(8,6)--(5,6)--(f2-3)--(t2-3)--(f2-1)--(t2-2);
\draw (f1-2)--(1,12.5)--(1,17.5)--(t1-1)--(8,17.5)--(8,12)--(5,12)--(f1-3)--(t1-3)--(f1-1)--(t1-2);
\end{scope}

\begin{scope}[blue]
\draw (t3-1)--(c3-3)--(f3-1)--(c3-1)--(t3-1);
\draw (t3-2)--(t3-3)--(t3-4)--(c3-4)--(f3-4)--(f3-3)--(f3-2)--(c3-2)--(t3-2);
\draw (f3-2)--(f3-3);

\draw (t2-1)--(c2-3)--(f2-1)--(c2-1)--(t2-1);
\draw (t2-2)--(t2-3)--(t2-4)--(c2-4)--(f2-4)--(f2-3)--(f2-2)--(c2-2)--(t2-2);
\draw (f2-2)--(f2-3);

\draw (t1-1)--(c1-3)--(f1-1)--(c1-1)--(t1-1);
\draw (t1-2)--(t1-3)--(t1-4)--(c1-4)--(f1-4)--(f1-3)--(f1-2)--(c1-2)--(t1-2);
\draw (f1-2)--(f1-3);

\draw (1)--(2) (1)--(c1-1) (2)--(c1-2) (d8-3)--(3)--(d8-4);
\end{scope}

\begin{scope}[blue]
\draw (d8-1)--(r8)--(d8-3);
\draw (d8-4)--(f8z3)--(d8-2)--(f8z2)--(d8-4)--(f8z1)--(d8-2);

\draw (d7-1)--(r7)--(d7-3);
\draw (d7-4)--(f7z3)--(d7-2)--(f7z2)--(d7-4)--(f7z1)--(d7-2);

\draw (d6-1)--(r6)--(d6-3);
\draw (d6-4)--(f6z3)--(d6-2)--(f6z2)--(d6-4)--(f6z1)--(d6-2);

\draw (d5-1)--(r5)--(d5-3);
\draw (d5-4)--(f5z3)--(d5-2)--(f5z2)--(d5-4)--(f5z1)--(d5-2);

\draw (d4-1)--(r4)--(d4-3);
\draw (d4-4)--(f4z3)--(d4-2)--(f4z2)--(d4-4)--(f4z1)--(d4-2);

\draw (d3-1)--(r3)--(d3-3);
\draw (d3-4)--(f3z3)--(d3-2)--(f3z2)--(d3-4)--(f3z1)--(d3-2);

\draw (d2-1)--(r2)--(d2-3);
\draw (d2-4)--(f2z3)--(d2-2)--(f2z2)--(d2-4)--(f2z1)--(d2-2);

\draw (d1-1)--(r1)--(d1-3);
\draw (d1-4)--(f1z3)--(d1-2)--(f1z2)--(d1-4)--(f1z1)--(d1-2);

\draw (d1-3)--(d2-1) (d1-4)--(d2-2) (d2-3)--(d3-1) (d2-4)--(d3-2) (d3-3)--(d4-1) (d3-4)--(d4-2) (d4-3)--(d5-1) (d4-4)--(d5-2) (d5-3)--(d6-1) (d5-4)--(d6-2) (d6-3)--(d7-1) (d6-4)--(d7-2) (d7-3)--(d8-1) (d7-4)--(d8-2); 

\draw (c1-3)--(c2-1) (c1-4)--(c2-2) (c2-3)--(c3-1) (c2-4)--(c3-2) (c3-3)--(d1-1) (c3-4)--(d1-2); 

\end{scope}

\begin{scope}[red]

\draw (t1-1)--(f1z1) (t1-3)--(f1z1) (t2-1)--(f1z2) (t2-3)--(f1z2) (t3-1)--(f1z3) (t3-3)--(f1z3);

\draw (t1-1)--(f2z1) (t1-3)--(f2z1) (t2-1)--(f2z2) (t2-3)--(f2z2) (f3-1)--(f2z3) (f3-3)--(f2z3);

\draw (t1-1)--(f3z1) (t1-3)--(f3z1) (f2-1)--(f3z2) (f2-3)--(f3z2) (t3-1)--(f3z3) (t3-3)--(f3z3);

\draw (t1-1)--(f4z1) (t1-3)--(f4z1) (f2-1)--(f4z2) (f2-3)--(f4z2) (f3-1)--(f4z3) (f3-3)--(f4z3);

\draw (f1-1)--(f5z1) (f1-3)--(f5z1) (t2-1)--(f5z2) (t2-3)--(f5z2) (t3-1)--(f5z3) (t3-3)--(f5z3);

\draw (f1-1)--(f6z1) (f1-3)--(f6z1) (t2-1)--(f6z2) (t2-3)--(f6z2) (f3-1)--(f6z3) (f3-3)--(f6z3);

\draw (f1-1)--(f7z1) (f1-3)--(f7z1) (f2-1)--(f7z2) (f2-3)--(f7z2) (t3-1)--(f7z3) (t3-3)--(f7z3);  

\draw (f1-1)--(f8z1) (f1-3)--(f8z1) (f2-1)--(f8z2) (f2-3)--(f8z2) (f3-1)--(f8z3) (f3-3)--(f8z3); 

\end{scope}

	\end{tikzpicture}
\caption{Gadget version of graph for formula $(x_1+x_2+x_3)(x_1+x_2+\overline x_3)(x_1+\overline x_2+x_3)(x_1+\overline x_2+\overline x_3)(\overline x_1+x_2+x_3)(\overline x_1+x_2+\overline x_3)(\overline x_1+\overline x_2+x_3)(\overline x_1+\overline x_2+\overline x_3)$.}
\label{fig:overview}
\end{figure}
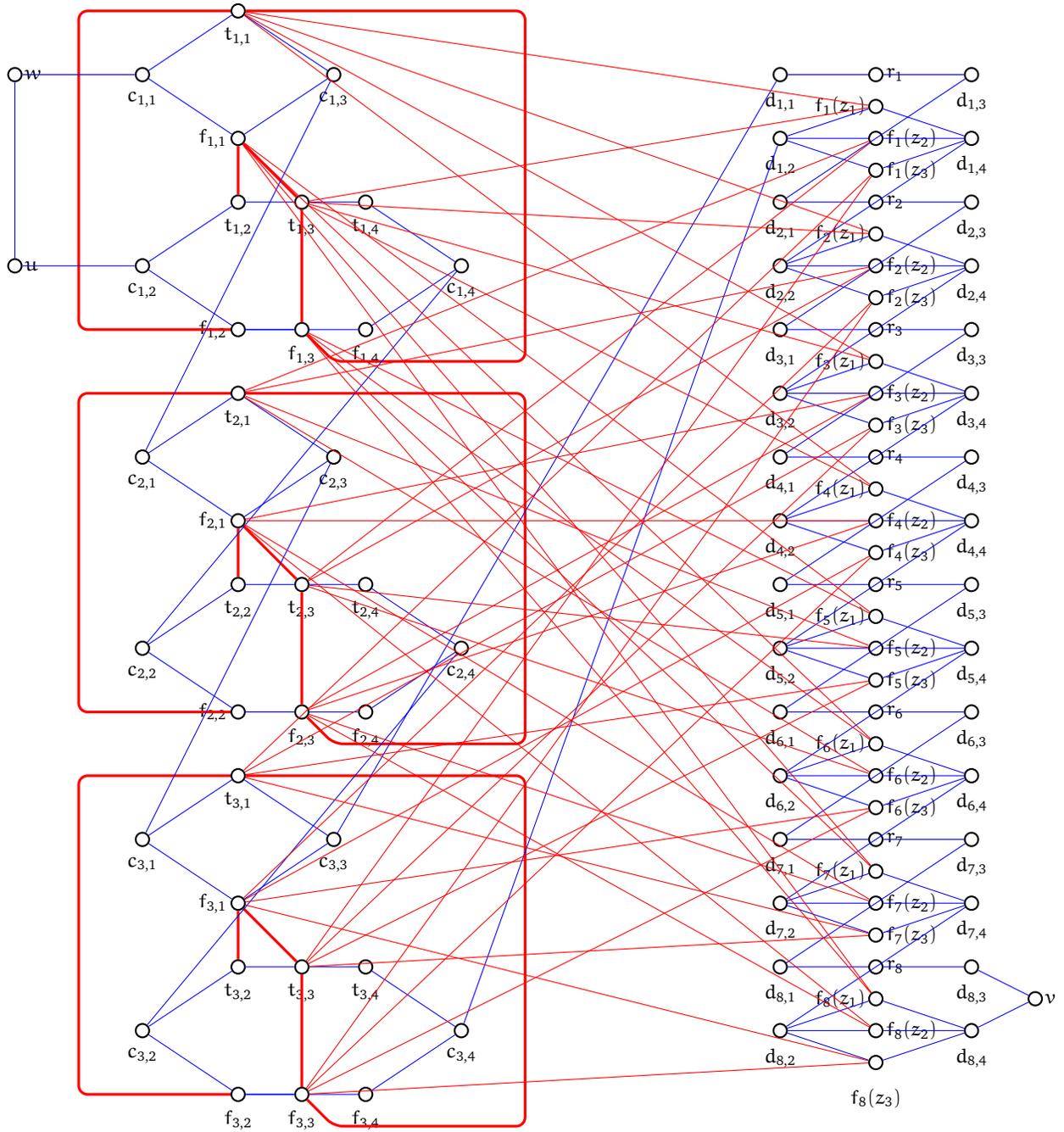

\begin{figure}[H]
	\centering\small
\begin{tikzpicture}[thick, scale=0.4]

\tikzstyle{place}=[circle,draw=black,fill=white,thick,
inner sep=0pt,minimum size=1mm]

\tikzstyle{every node}=[circle,draw=black,fill=white,thick,minimum size=3pt,
inner sep=0pt]

\draw (8,0) node (v) [label=right:$v$] {};
\draw (2,3) node (r8) [label=right:$r_8$] {};
\draw (2,4) node (d8-1) [label=right:$d_{8,1}$] {};
\draw (14,4) node (d8-2) [label=right:$d_{8,2}$] {};
\draw (2,2) node (d8-3) [label=right:$d_{8,3}$] {};
\draw (14,2) node (d8-4) [label=right:$d_{8,4}$] {}; 
\draw (13,3) node (f8z1) [label=left:$f_8({z}_1)$] {}; 
\node(f8z2)[label={[label distance=0.3cm]right:{$f_8({z}_2)$}}] at (14,3) {};
\node(f8z3)[label={[label distance=0.9cm]right:{$f_8({z}_3)$}}] at (15,3) {};

\draw (2,6) node (r7) [label=right:$r_7$] {};
\draw (2,7) node (d7-1) [label=right:$d_{7,1}$] {};
\draw (14,7) node (d7-2) [label=right:$d_{7,2}$] {};
\draw (2,5) node (d7-3) [label=right:$d_{7,3}$] {};
\draw (14,5) node (d7-4) [label=right:$d_{7,4}$] {};
\draw (13,6) node (f7z1) [label=left:$f_7({z}_1)$] {};
\node(f7z2)[label={[label distance=0.3cm]right:{$f_7({z}_2)$}}] at (14,6) {};
\node(f7z3)[label={[label distance=0.9cm]right:{$f_7({z}_3)$}}] at (15,6) {};

\draw (2,9) node (r6) [label=right:$r_6$] {};
\draw (2,10) node (d6-1) [label=right:$d_{6,1}$] {};
\draw (14,10) node (d6-2) [label=right:$d_{6,2}$] {};
\draw (2,8) node (d6-3) [label=right:$d_{6,3}$] {};
\draw (14,8) node (d6-4) [label=right:$d_{6,4}$] {}; 
\draw (13,9) node (f6z1) [label=left:$f_6({z}_1)$] {};
\node(f6z2)[label={[label distance=0.3cm]right:{$f_6({z}_2)$}}] at (14,9) {};
\node(f6z3)[label={[label distance=0.9cm]right:{$f_6({z}_3)$}}] at (15,9) {};

\draw (2,12) node (r5) [label=right:$r_5$] {};
\draw (2,13) node (d5-1) [label=right:$d_{5,1}$] {};
\draw (14,13) node (d5-2) [label=right:$d_{5,2}$] {};
\draw (2,11) node (d5-3) [label=right:$d_{5,3}$] {};
\draw (14,11) node (d5-4) [label=right:$d_{5,4}$] {}; 
\draw (13,12) node (f5z1) [label=left:$f_5({z}_1)$] {}; 
\node(f5z2)[label={[label distance=0.3cm]right:{$f_5({z}_2)$}}] at (14,12) {};
\node(f5z3)[label={[label distance=0.9cm]right:{$f_5({z}_3)$}}] at (15,12) {};

\draw (2,15) node (r4) [label=right:$r_4$] {};
\draw (2,16) node (d4-1) [label=right:$d_{4,1}$] {};
\draw (14,16) node (d4-2) [label=right:$d_{4,2}$] {};
\draw (2,14) node (d4-3) [label=right:$d_{4,3}$] {};
\draw (14,14) node (d4-4) [label=right:$d_{4,4}$] {}; 
\draw (13,15) node (f4z1) [label=left:$f_4({z}_1)$] {}; 
\node(f4z2)[label={[label distance=0.3cm]right:{$f_4({z}_2)$}}] at (14,15) {};
\node(f4z3)[label={[label distance=0.9cm]right:{$f_4({z}_3)$}}] at (15,15) {};

\draw (2,18) node (r3) [label=right:$r_3$] {};
\draw (2,19) node (d3-1) [label=right:$d_{3,1}$] {};
\draw (14,19) node (d3-2) [label=right:$d_{3,2}$] {};
\draw (2,17) node (d3-3) [label=right:$d_{3,3}$] {};
\draw (14,17) node (d3-4) [label=right:$d_{3,4}$] {}; 
\draw (13,18) node (f3z1) [label=left:$f_3({z}_1)$] {}; 
\node(f3z2)[label={[label distance=0.3cm]right:{$f_3({z}_2)$}}] at (14,18) {};
\node(f3z3)[label={[label distance=0.9cm]right:{$f_3({z}_3)$}}] at (15,18) {};

\draw (2,21) node (r2) [label=right:$r_2$] {};
\draw (2,22) node (d2-1) [label=right:$d_{2,1}$] {};
\draw (14,22) node (d2-2) [label=right:$d_{2,2}$] {};
\draw (2,20) node (d2-3) [label=right:$d_{2,3}$] {};
\draw (14,20) node (d2-4) [label=right:$d_{2,4}$] {}; 
\draw (13,21) node (f2z1) [label=left:$f_2({z}_1)$] {}; 
\node(f2z2)[label={[label distance=0.3cm]right:{$f_2({z}_2)$}}] at (14,21) {};
\node(f2z3)[label={[label distance=0.9cm]right:{$f_2({z}_3)$}}] at (15,21) {};

\draw (2,24) node (r1) [label=right:$r_1$] {};
\draw (2,25) node (d1-1) [label=right:$d_{1,1}$] {};
\draw (14,25) node (d1-2) [label=right:$d_{1,2}$] {};
\draw (2,23) node (d1-3) [label=right:$d_{1,3}$] {};
\draw (14,23) node (d1-4) [label=right:$d_{1,4}$] {}; 
\draw (13,24) node (f1z1) [label=left:$f_1({z}_1)$] {}; 
\node(f1z2)[label={[label distance=0.3cm]right:{$f_1({z}_2)$}}] at (14,24) {};
\node(f1z3)[label={[label distance=0.9cm]right:{$f_1({z}_3)$}}] at (15,24) {};

\draw (2,31) node (c3-1) [label=right:$c_{3,1}$] {};
\draw (1,28.5) node (f3-1) [label=left:$f_{3,1}$] {};
\draw (3,29.5) node (t3-1) [label=right:$t_{3,1}$] {};
\draw (2,27) node (c3-3) [label=right:$c_{3,3}$] {};

\draw (14,31) node (c3-2) [label=right:$c_{3,2}$] {};
\draw (13,30) node (f3-2) [label=left:$f_{3,2}$] {};
\draw (13,29) node (f3-3) [label=left:$f_{3,3}$] {};
\draw (13,28) node (f3-4) [label=left:$f_{3,4}$] {};
\draw (15,30) node (t3-2) [label=right:$t_{3,2}$] {};
\draw (15,29) node (t3-3) [label=right:$t_{3,3}$] {};
\draw (15,28) node (t3-4) [label=right:$t_{3,4}$] {};
\draw (14,27) node (c3-4) [label=right:$c_{3,4}$] {};

\draw (2,36) node (c2-1) [label=right:$c_{2,1}$] {};
\draw (1,33.5) node (f2-1) [label=left:$f_{2,1}$] {};
\draw (3,34.5) node (t2-1) [label=right:$t_{2,1}$] {};
\draw (2,32) node (c2-3) [label=right:$c_{2,3}$] {};

\draw (14,36) node (c2-2) [label=right:$c_{2,2}$] {};
\draw (13,35) node (f2-2) [label=left:$f_{2,2}$] {};
\draw (13,34) node (f2-3) [label=left:$f_{2,3}$] {};
\draw (13,33) node (f2-4) [label=left:$f_{2,4}$] {};
\draw (15,35) node (t2-2) [label=right:$t_{2,2}$] {};
\draw (15,34) node (t2-3) [label=right:$t_{2,3}$] {};
\draw (15,33) node (t2-4) [label=right:$t_{2,4}$] {};
\draw (14,32) node (c2-4) [label=right:$c_{2,4}$] {};

\draw (2,41) node (c1-1) [label=right:$c_{1,1}$] {};
\draw (1,38.5) node (f1-1) [label=left:$f_{1,1}$] {};
\draw (3,39.5) node (t1-1) [label=right:$t_{1,1}$] {};
\draw (2,37) node (c1-3) [label=right:$c_{1,3}$] {};

\draw (14,41) node (c1-2) [label=right:$c_{1,2}$] {};
\draw (13,40) node (f1-2) [label=left:$f_{1,2}$] {};
\draw (13,39) node (f1-3) [label=left:$f_{1,3}$] {};
\draw (13,38) node (f1-4) [label=left:$f_{1,4}$] {};
\draw (15,40) node (t1-2) [label=right:$t_{1,2}$] {};
\draw (15,39) node (t1-3) [label=right:$t_{1,3}$] {};
\draw (15,38) node (t1-4) [label=right:$t_{1,4}$] {};
\draw (14,37) node (c1-4) [label=right:$c_{1,4}$] {};

\draw (6,43) node (w) [label=left:$w$] {};
\draw (10,43) node (u) [label=right:$u$] {};

\begin{scope}[red]
\draw (f3-2)--(t3-1)--(f3-3)--(t3-3)--(f3-1)--(t3-2);
\draw (f2-2)--(t2-1)--(f2-3)--(t2-3)--(f2-1)--(t2-2);
\draw (f1-2)--(t1-1)--(f1-3)--(t1-3)--(f1-1)--(t1-2);

\draw (t1-1)--(f1z1) (t1-3)--(f1z1) (t2-1)--(f1z2) (t2-3)--(f1z2) (t3-1)--(f1z3) (t3-3)--(f1z3);

\draw (t1-1)--(f2z1) (t1-3)--(f2z1) (t2-1)--(f2z2) (t2-3)--(f2z2) (f3-1)--(f2z3) (f3-3)--(f2z3);

\draw (t1-1)--(f3z1) (t1-3)--(f3z1) (f2-1)--(f3z2) (f2-3)--(f3z2) (t3-1)--(f3z3) (t3-3)--(f3z3);

\draw (t1-1)--(f4z1) (t1-3)--(f4z1) (f2-1)--(f4z2) (f2-3)--(f4z2) (f3-1)--(f4z3) (f3-3)--(f4z3);

\draw (f1-1)--(f5z1) (f1-3)--(f5z1) (t2-1)--(f5z2) (t2-3)--(f5z2) (t3-1)--(f5z3) (t3-3)--(f5z3);

\draw (f1-1)--(f6z1) (f1-3)--(f6z1) (t2-1)--(f6z2) (t2-3)--(f6z2) (f3-1)--(f6z3) (f3-3)--(f6z3);

\draw (f1-1)--(f7z1) (f1-3)--(f7z1) (f2-1)--(f7z2) (f2-3)--(f7z2) (t3-1)--(f7z3) (t3-3)--(f7z3);  

\draw (f1-1)--(f8z1) (f1-3)--(f8z1) (f2-1)--(f8z2) (f2-3)--(f8z2) (f3-1)--(f8z3) (f3-3)--(f8z3); 

\end{scope}

\begin{scope}[blue]
\draw (t3-1)--(c3-3)--(f3-1)--(c3-1)--(t3-1);
\draw (t3-2)--(t3-3)--(t3-4)--(c3-4)--(f3-4)--(f3-3)--(f3-2)--(c3-2)--(t3-2);
\draw (f3-2)--(f3-3);

\draw (t2-1)--(c2-3)--(f2-1)--(c2-1)--(t2-1);
\draw (t2-2)--(t2-3)--(t2-4)--(c2-4)--(f2-4)--(f2-3)--(f2-2)--(c2-2)--(t2-2);
\draw (f2-2)--(f2-3);

\draw (t1-1)--(c1-3)--(f1-1)--(c1-1)--(t1-1);
\draw (t1-2)--(t1-3)--(t1-4)--(c1-4)--(f1-4)--(f1-3)--(f1-2)--(c1-2)--(t1-2);
\draw (f1-2)--(f1-3);
\end{scope}

\begin{scope}[blue]
\draw (d1-1)--(r1)--(d1-3)--(d2-1)--(r2)--(d2-3)--(d3-1)--(r3)--(d3-3)--(d4-1)--(r4)--(d4-3)--(d5-1)--(r5)--(d5-3)--(d6-1)--(r6)--(d6-3)--(d7-1)--(r7)--(d7-3)--(d8-1)--(r8)--(d8-3);

\draw (d8-4)--(f8z3)--(d8-2)--(f8z2)--(d8-4)--(f8z1)--(d8-2);

\draw (d7-4)--(f7z3)--(d7-2)--(f7z2)--(d7-4)--(f7z1)--(d7-2);

\draw (d6-4)--(f6z3)--(d6-2)--(f6z2)--(d6-4)--(f6z1)--(d6-2);

\draw (d5-4)--(f5z3)--(d5-2)--(f5z2)--(d5-4)--(f5z1)--(d5-2);

\draw (d4-4)--(f4z3)--(d4-2)--(f4z2)--(d4-4)--(f4z1)--(d4-2);

\draw (d3-4)--(f3z3)--(d3-2)--(f3z2)--(d3-4)--(f3z1)--(d3-2);

\draw (d2-4)--(f2z3)--(d2-2)--(f2z2)--(d2-4)--(f2z1)--(d2-2);

\draw (d1-4)--(f1z3)--(d1-2)--(f1z2)--(d1-4)--(f1z1)--(d1-2);

\draw (d1-3)--(d2-1) (d1-4)--(d2-2) (d2-3)--(d3-1) (d2-4)--(d3-2); 
\draw (d3-4)--(d4-2) (d4-4)--(d5-2) (d5-4)--(d6-2) (d6-4)--(d7-2) (d7-4)--(d8-2);
\draw (c1-3)--(c2-1) (c1-4)--(c2-2) (c2-3)--(c3-1) (c2-4)--(c3-2); 
\draw (c3-3)--(d1-1);
\draw (c3-4)--(d1-2);
\draw (d8-3)--(v)--(d8-4);
\draw (c1-1)--(w)--(u)--(c1-2);

\end{scope}

\end{tikzpicture}

\caption{Untwisted version of graph for formula $(x_1+x_2+x_3)(x_1+x_2+\overline x_3)(x_1+\overline x_2+x_3)(x_1+\overline x_2+\overline x_3)(\overline x_1+x_2+x_3)(\overline x_1+x_2+\overline x_3)(\overline x_1+\overline x_2+x_3)(\overline x_1+\overline x_2+\overline x_3)$.}
\label{fig:overview}
\end{figure}
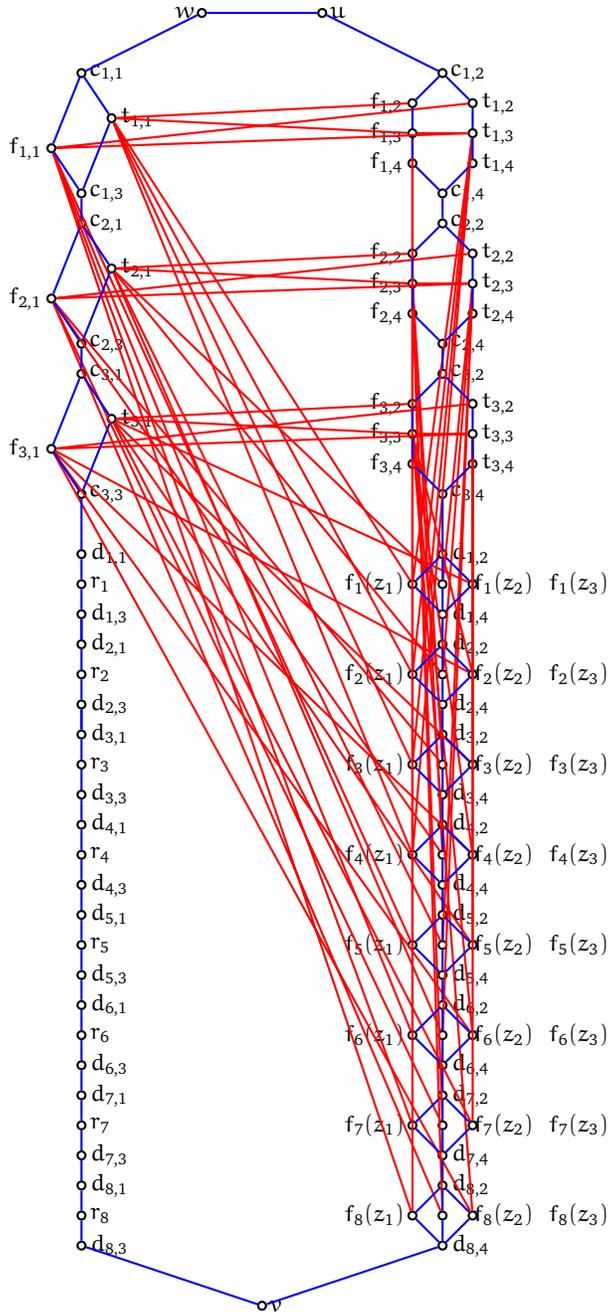

\begin{figure}[H]
	\centering\small
	
	\begin{tikzpicture}[thick, scale=0.4]
	
	\tikzstyle{place}=[circle,draw=black,fill=white,thick,
	inner sep=0pt,minimum size=1mm]
	
	\tikzstyle{every node}=[circle,draw=black,fill=white,thick,minimum size=3pt,
	inner sep=0pt]

	\draw (8,0) node (v) [label=right:$v$] {};
	\draw (2,3) node (r8) [label=right:$r_8$] {};
	\draw (2,4) node (d8-1) [label=right:$d_{8,1}$] {};
	\draw (14,4) node (d8-2) [label=right:$d_{8,2}$] {};
	\draw (2,2) node (d8-3) [label=right:$d_{8,3}$] {};
	\draw (14,2) node (d8-4) [label=right:$d_{8,4}$] {}; 
	\draw (13,3) node (f8z1) [label=left:$f_8({z}_1)$] {}; 
	\node(f8z2)[label={[label distance=0.3cm]right:{$f_8({z}_2)$}}] at (14,3) {};
	\node(f8z3)[label={[label distance=0.9cm]right:{$f_8({z}_3)$}}] at (15,3) {};

	\draw (2,6) node (r7) [label=right:$r_7$] {};
	\draw (2,7) node (d7-1) [label=right:$d_{7,1}$] {};
	\draw (14,7) node (d7-2) [label=right:$d_{7,2}$] {};
	\draw (2,5) node (d7-3) [label=right:$d_{7,3}$] {};
	\draw (14,5) node (d7-4) [label=right:$d_{7,4}$] {};
	\draw (13,6) node (f7z1) [label=left:$f_7({z}_1)$] {};
	\node(f7z2)[label={[label distance=0.3cm]right:{$f_7({z}_2)$}}] at (14,6) {};
	\node(f7z3)[label={[label distance=0.9cm]right:{$f_7({z}_3)$}}] at (15,6) {};

	\draw (2,9) node (r6) [label=right:$r_6$] {};
	\draw (2,10) node (d6-1) [label=right:$d_{6,1}$] {};
	\draw (14,10) node (d6-2) [label=right:$d_{6,2}$] {};
	\draw (2,8) node (d6-3) [label=right:$d_{6,3}$] {};
	\draw (14,8) node (d6-4) [label=right:$d_{6,4}$] {}; 
	\draw (13,9) node (f6z1) [label=left:$f_6({z}_1)$] {};
	\node(f6z2)[label={[label distance=0.3cm]right:{$f_6({z}_2)$}}] at (14,9) {};
	\node(f6z3)[label={[label distance=0.9cm]right:{$f_6({z}_3)$}}] at (15,9) {};

	\draw (2,12) node (r5) [label=right:$r_5$] {};
	\draw (2,13) node (d5-1) [label=right:$d_{5,1}$] {};
	\draw (14,13) node (d5-2) [label=right:$d_{5,2}$] {};
	\draw (2,11) node (d5-3) [label=right:$d_{5,3}$] {};
	\draw (14,11) node (d5-4) [label=right:$d_{5,4}$] {}; 
	\draw (13,12) node (f5z1) [label=left:$f_5({z}_1)$] {}; 
	\node(f5z2)[label={[label distance=0.3cm]right:{$f_5({z}_2)$}}] at (14,12) {};
	\node(f5z3)[label={[label distance=0.9cm]right:{$f_5({z}_3)$}}] at (15,12) {};

	\draw (2,15) node (r4) [label=right:$r_4$] {};
	\draw (2,16) node (d4-1) [label=right:$d_{4,1}$] {};
	\draw (14,16) node (d4-2) [label=right:$d_{4,2}$] {};
	\draw (2,14) node (d4-3) [label=right:$d_{4,3}$] {};
	\draw (14,14) node (d4-4) [label=right:$d_{4,4}$] {}; 
	\draw (13,15) node (f4z1) [label=left:$f_4({z}_1)$] {}; 
	\node(f4z2)[label={[label distance=0.3cm]right:{$f_4({z}_2)$}}] at (14,15) {};
	\node(f4z3)[label={[label distance=0.9cm]right:{$f_4({z}_3)$}}] at (15,15) {};

	\draw (2,18) node (r3) [label=right:$r_3$] {};
	\draw (2,19) node (d3-1) [label=right:$d_{3,1}$] {};
	\draw (14,19) node (d3-2) [label=right:$d_{3,2}$] {};
	\draw (2,17) node (d3-3) [label=right:$d_{3,3}$] {};
	\draw (14,17) node (d3-4) [label=right:$d_{3,4}$] {}; 
	\draw (13,18) node (f3z1) [label=left:$f_3({z}_1)$] {}; 
	\node(f3z2)[label={[label distance=0.3cm]right:{$f_3({z}_2)$}}] at (14,18) {};
	\node(f3z3)[label={[label distance=0.9cm]right:{$f_3({z}_3)$}}] at (15,18) {};

	\draw (2,21) node (r2) [label=right:$r_2$] {};
	\draw (2,22) node (d2-1) [label=right:$d_{2,1}$] {};
	\draw (14,22) node (d2-2) [label=right:$d_{2,2}$] {};
	\draw (2,20) node (d2-3) [label=right:$d_{2,3}$] {};
	\draw (14,20) node (d2-4) [label=right:$d_{2,4}$] {}; 
	\draw (13,21) node (f2z1) [label=left:$f_2({z}_1)$] {}; 
	\node(f2z2)[label={[label distance=0.3cm]right:{$f_2({z}_2)$}}] at (14,21) {};
	\node(f2z3)[label={[label distance=0.9cm]right:{$f_2({z}_3)$}}] at (15,21) {};

	\draw (2,24) node (r1) [label=right:$r_1$] {};
	\draw (2,25) node (d1-1) [label=right:$d_{1,1}$] {};
	\draw (14,25) node (d1-2) [label=right:$d_{1,2}$] {};
	\draw (2,23) node (d1-3) [label=right:$d_{1,3}$] {};
	\draw (14,23) node (d1-4) [label=right:$d_{1,4}$] {}; 
	\draw (13,24) node (f1z1) [label=left:$f_1({z}_1)$] {}; 
	\node(f1z2)[label={[label distance=0.3cm]right:{$f_1({z}_2)$}}] at (14,24) {};
	\node(f1z3)[label={[label distance=0.9cm]right:{$f_1({z}_3)$}}] at (15,24) {};

	\draw (2,31) node (c3-1) [label=right:$c_{3,1}$] {};
	\draw (1,28.5) node (f3-1) [label=left:$f_{3,1}$] {};
	\draw (3,29.5) node (t3-1) [label=right:$t_{3,1}$] {};
	\draw (2,27) node (c3-3) [label=right:$c_{3,3}$] {};

	\draw (14,31) node (c3-2) [label=right:$c_{3,2}$] {};
	\draw (13,30) node (f3-2) [label=left:$f_{3,2}$] {};
	\draw (13,29) node (f3-3) [label=left:$f_{3,3}$] {};
	\draw (13,28) node (f3-4) [label=left:$f_{3,4}$] {};
	\draw (15,30) node (t3-2) [label=right:$t_{3,2}$] {};
	\draw (15,29) node (t3-3) [label=right:$t_{3,3}$] {};
	\draw (15,28) node (t3-4) [label=right:$t_{3,4}$] {};
	\draw (14,27) node (c3-4) [label=right:$c_{3,4}$] {};

	\draw (2,36) node (c2-1) [label=right:$c_{2,1}$] {};
	\draw (1,33.5) node (f2-1) [label=left:$f_{2,1}$] {};
	\draw (3,34.5) node (t2-1) [label=right:$t_{2,1}$] {};
	\draw (2,32) node (c2-3) [label=right:$c_{2,3}$] {};

	\draw (14,36) node (c2-2) [label=right:$c_{2,2}$] {};
	\draw (13,35) node (f2-2) [label=left:$f_{2,2}$] {};
	\draw (13,34) node (f2-3) [label=left:$f_{2,3}$] {};
	\draw (13,33) node (f2-4) [label=left:$f_{2,4}$] {};
	\draw (15,35) node (t2-2) [label=right:$t_{2,2}$] {};
	\draw (15,34) node (t2-3) [label=right:$t_{2,3}$] {};
	\draw (15,33) node (t2-4) [label=right:$t_{2,4}$] {};
	\draw (14,32) node (c2-4) [label=right:$c_{2,4}$] {};

	\draw (2,41) node (c1-1) [label=right:$c_{1,1}$] {};
	\draw (1,38.5) node (f1-1) [label=left:$f_{1,1}$] {};
	\draw (3,39.5) node (t1-1) [label=right:$t_{1,1}$] {};
	\draw (2,37) node (c1-3) [label=right:$c_{1,3}$] {};

	\draw (14,41) node (c1-2) [label=right:$c_{1,2}$] {};
	\draw (13,40) node (f1-2) [label=left:$f_{1,2}$] {};
	\draw (13,39) node (f1-3) [label=left:$f_{1,3}$] {};
	\draw (13,38) node (f1-4) [label=left:$f_{1,4}$] {};
	\draw (15,40) node (t1-2) [label=right:$t_{1,2}$] {};
	\draw (15,39) node (t1-3) [label=right:$t_{1,3}$] {};
	\draw (15,38) node (t1-4) [label=right:$t_{1,4}$] {};
	\draw (14,37) node (c1-4) [label=right:$c_{1,4}$] {};
	
	\draw (6,43) node (w) [label=left:$w$] {};
	\draw (10,43) node (u) [label=right:$u$] {};

	
	\begin{scope}[blue]
	\draw (t3-1)--(c3-3)--(f3-1)--(c3-1)--(t3-1);
	\draw (t3-2)--(t3-3)--(t3-4)--(c3-4)--(f3-4)--(f3-3)--(f3-2)--(c3-2)--(t3-2);
	\draw (f3-2)--(f3-3);
	
	\draw (t2-1)--(c2-3)--(f2-1)--(c2-1)--(t2-1);
	\draw (t2-2)--(t2-3)--(t2-4)--(c2-4)--(f2-4)--(f2-3)--(f2-2)--(c2-2)--(t2-2);
	\draw (f2-2)--(f2-3);
	
	\draw (t1-1)--(c1-3)--(f1-1)--(c1-1)--(t1-1);
	\draw (t1-2)--(t1-3)--(t1-4)--(c1-4)--(f1-4)--(f1-3)--(f1-2)--(c1-2)--(t1-2);
	\draw (f1-2)--(f1-3);
	\end{scope}
	
	\begin{scope}[blue]
	\draw (d1-1)--(r1)--(d1-3)--(d2-1)--(r2)--(d2-3)--(d3-1)--(r3)--(d3-3)--(d4-1)--(r4)--(d4-3)--(d5-1)--(r5)--(d5-3)--(d6-1)--(r6)--(d6-3)--(d7-1)--(r7)--(d7-3)--(d8-1)--(r8)--(d8-3);
	
	\draw (d8-4)--(f8z3)--(d8-2)--(f8z2)--(d8-4)--(f8z1)--(d8-2);

	\draw (d7-4)--(f7z3)--(d7-2)--(f7z2)--(d7-4)--(f7z1)--(d7-2);

	\draw (d6-4)--(f6z3)--(d6-2)--(f6z2)--(d6-4)--(f6z1)--(d6-2);

	\draw (d5-4)--(f5z3)--(d5-2)--(f5z2)--(d5-4)--(f5z1)--(d5-2);

	\draw (d4-4)--(f4z3)--(d4-2)--(f4z2)--(d4-4)--(f4z1)--(d4-2);

	\draw (d3-4)--(f3z3)--(d3-2)--(f3z2)--(d3-4)--(f3z1)--(d3-2);

	\draw (d2-4)--(f2z3)--(d2-2)--(f2z2)--(d2-4)--(f2z1)--(d2-2);

	\draw (d1-4)--(f1z3)--(d1-2)--(f1z2)--(d1-4)--(f1z1)--(d1-2);
	
	\draw (d1-3)--(d2-1) (d1-4)--(d2-2) (d2-3)--(d3-1) (d2-4)--(d3-2); 
	\draw (d3-4)--(d4-2) (d4-4)--(d5-2) (d5-4)--(d6-2) (d6-4)--(d7-2) (d7-4)--(d8-2);
	\draw (c1-3)--(c2-1) (c1-4)--(c2-2) (c2-3)--(c3-1) (c2-4)--(c3-2); 
	\draw (c3-3)--(d1-1);
	\draw (c3-4)--(d1-2);
	\draw (d8-3)--(v)--(d8-4);
	\draw (c1-1)--(w)--(u)--(c1-2);
	
	\end{scope}
	
	\end{tikzpicture}
	
	\caption{Constraint free version of graph for formula $(x_1+x_2+x_3)(x_1+x_2+\overline x_3)(x_1+\overline x_2+x_3)(x_1+\overline x_2+\overline x_3)(\overline x_1+x_2+x_3)(\overline x_1+x_2+\overline x_3)(\overline x_1+\overline x_2+x_3)(\overline x_1+\overline x_2+\overline x_3)$.}
	\label{fig:overview}
\end{figure}
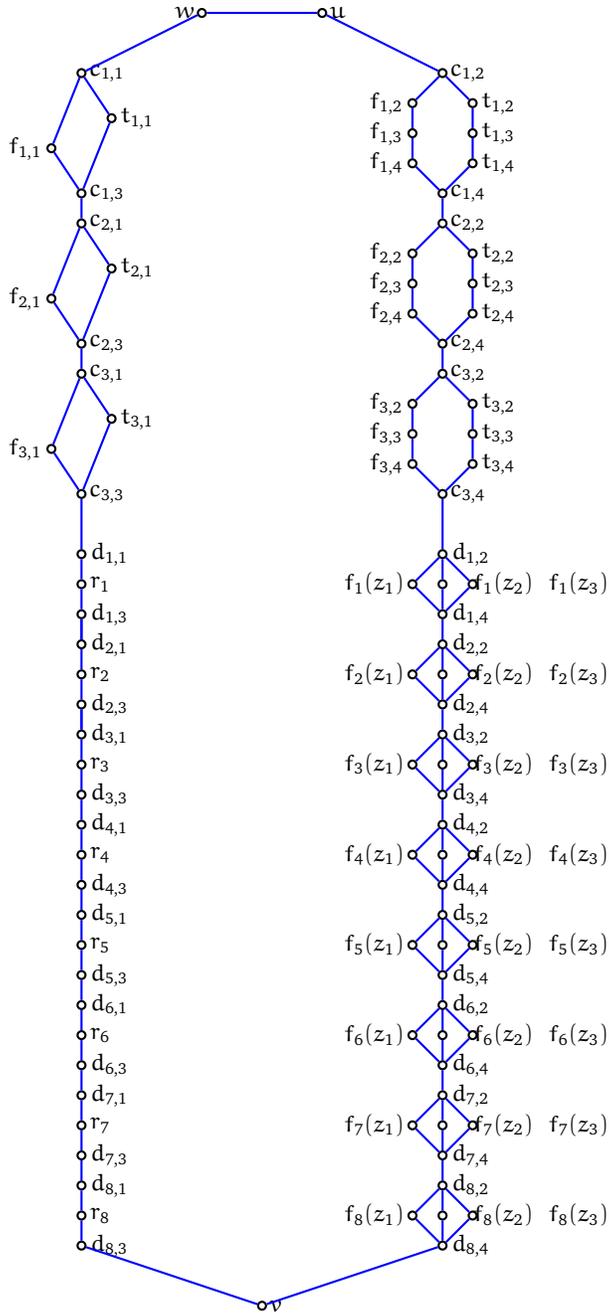

\begin{figure}[H]
	\centering\small
	\begin{tikzpicture}[thick, scale=0.8]
	
	\tikzstyle{every node}=[circle,draw=black,fill=white,thick,minimum size=8pt,
	inner sep=0pt]
	\draw (2,0) node (1)  {};
	\draw (5,.5) node (2)  {};
	\draw (7,0) node (3)  {};
	\draw (1,1) node (4)  {};
	\draw (6.5,1.5) node (5)  {};
	\draw (7.5,1) node (5b)  {};
	\draw (9,1.5) node (6)  {};
	\draw (0,2.5) node (7)  {};
	\draw (6.5,2.5) node (8)  {};
	\draw (7.5,2.5) node (9)  {};
	\draw (1,3.5) node (10)  {};
	\draw (0,3.5) node (11)  {};
	\draw (5.5,4) node (12)  {};
	\draw (8,3.5) node (13)  {};
	\draw (3,4.5) node (14)  {};
	\draw (6.5,4) node (15)  {};
	\draw (1,4.5) node (16)  {};
	\draw (4,4.5) node (17)  {};
	\draw (7.5,4.5) node (18)  {};
	\draw (2,5.5) node (19)  {};
	\draw (3.5,6) node (20)  {};
	\draw (6,6) node (21)  {};
	\draw (4.5,6.5) node (22)  {};
	\draw (9,4) node (23)  {};
	\draw (10,0.5) node (24)  {};

	\draw (2)--(1)--(3)--(6)--(9)--(13)--(18)--(21)--(22);
	\draw (2)--(5)--(8)--(12)--(17);
	\draw (5)--(5b)--(6);
	\draw (1)--(4)--(7)--(10)--(14)--(17);
	\draw (7)--(11)--(16)--(19)--(20)--(22);
	\draw (9)--(15)--(18);
	\draw (14)--(19) (9)--(23)--(18);
	\draw (8)--(24)--(4);

	\end{tikzpicture}
	\caption{A hole complex.}
	\label{fig:overview}
\end{figure}
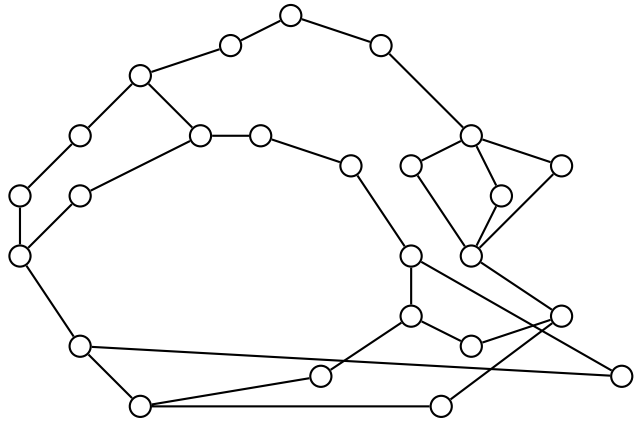

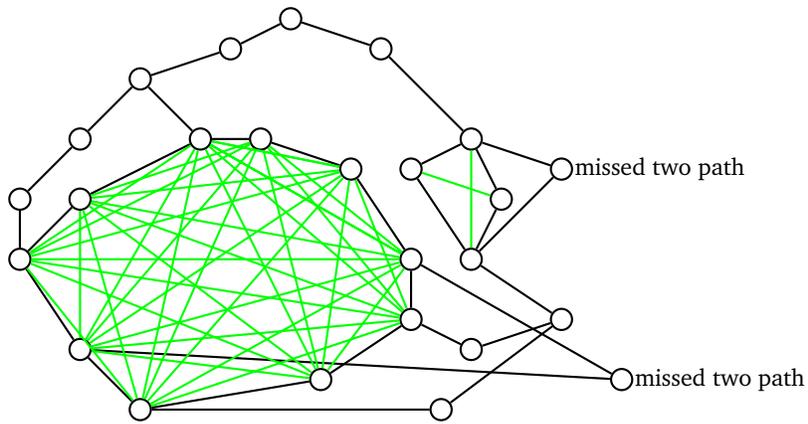
\begin{figure}[H]
	\centering\small
	
	\begin{tikzpicture}[thick, scale=0.8]
	
	\tikzstyle{every node}=[circle,draw=black,fill=white,thick,minimum size=8pt,
	inner sep=0pt]
	\draw (2,0) node (1)  {};
	\draw (5,.5) node (2)  {};
	\draw (7,0) node (3)  {};
	\draw (1,1) node (4)  {};
	\draw (6.5,1.5) node (5)  {};
	\draw (7.5,1) node (5b)  {};
	\draw (9,1.5) node (6)  {};
	\draw (0,2.5) node (7)  {};
	\draw (6.5,2.5) node (8)  {};
	\draw (7.5,2.5) node (9)  {};
	\draw (1,3.5) node (10)  {};
	\draw (0,3.5) node (11)  {};
	\draw (5.5,4) node (12)  {};
	\draw (8,3.5) node (13)  {};
	\draw (3,4.5) node (14)  {};
	\draw (6.5,4) node (15)  {};
	\draw (1,4.5) node (16)  {};
	\draw (4,4.5) node (17)  {};
	\draw (7.5,4.5) node (18)  {}; 
	\draw (2,5.5) node (19)  {};
	\draw (3.5,6) node (20)  {};
	\draw (6,6) node (21)  {};
	\draw (4.5,6.5) node (22)  {};
	\draw (9,4) node (23)  [label=right:missed two path] {};
	\draw (10,0.5) node (24) [label=right:missed two path] {};
	
	\draw (2)--(1)-- (3)--(6)--(9)--(13)--(18)--(21)--(22);
	\draw (2)--(5)--(8)--(12)--(17);
	\draw (5)--(5b)--(6);
	\draw (1)--(4)--(7)--(10)--(14)--(17);
	\draw (7)--(11)--(16)--(19)--(20)--(22);
	\draw (9)--(15)--(18);
	\draw (14)--(19) (9)--(23)--(18);
	\draw (8)--(24)--(4);
	
	\begin{scope}[green]
	\draw (1)--(5)--(12)--(14)--(7)--(1)--(8)--(17)--(10)--(4)--(2)--(8)--(14)--(4)--(5)--(17)--(7)--(2)--(12)--(10)--(1)--(12)--(4)--(8)--(14)--(1)--(17)--(4);
	\draw (17)--(2)--(10)--(5);
	\draw (2)--(14)--(12);
	\draw (8)--(10);
	\draw (8)--(7)--(12);  
	\draw (7)--(5);
	\draw (14)--(5);
	\draw (9)--(18);
	\draw (15)--(13);
	
	\end{scope}
	
	\end{tikzpicture}
	
	\caption{Partially filled hole complex showing missed two paths.}
	\label{fig:overview}
\end{figure}

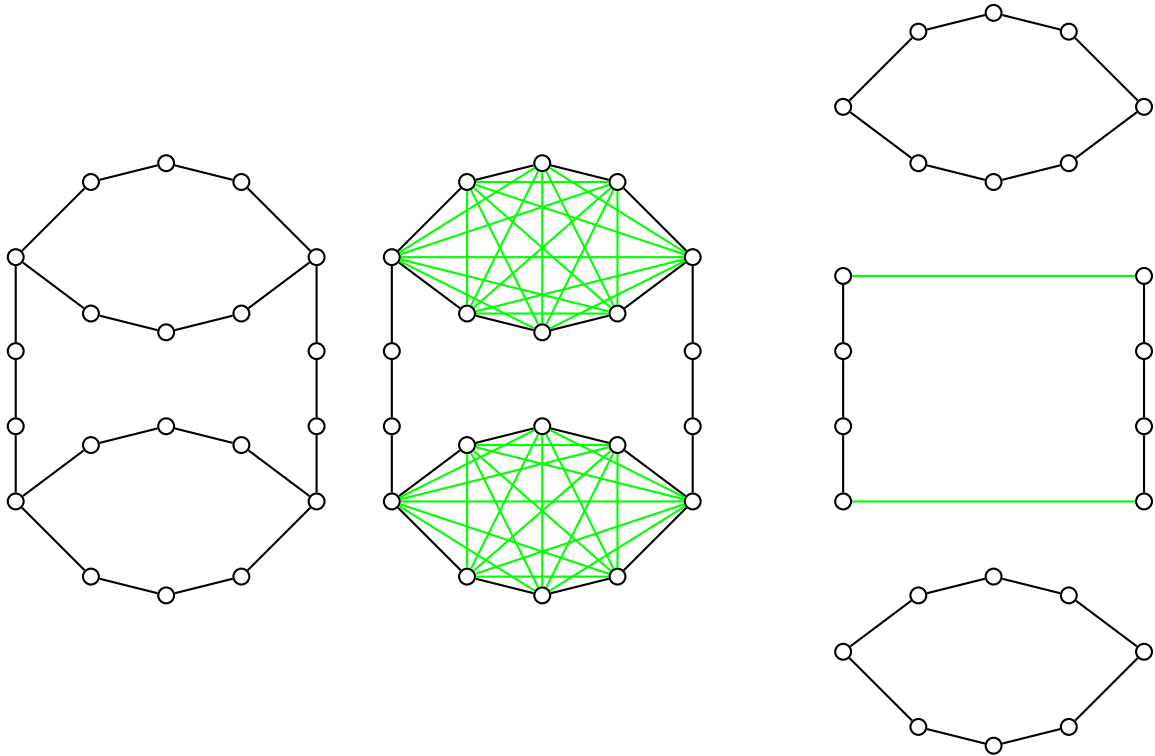
\begin{figure}[H]
	\centering\small
	\begin{tikzpicture}[thick, scale=0.5]
	
	\tikzstyle{every node}=[circle,draw=black,fill=white,thick,minimum size=6pt,
	inner sep=0pt]
	\draw (0,20) node (1) {};
	\draw (2,22) node (2) {};
	\draw (4,22.5) node (3) {};
	\draw (6,22) node (4) {};
	\draw (8,20) node (5) {};
	\draw (2,18.5) node (6) {};
	\draw (4,18) node (7) {};
	\draw (6,18.5) node (8) {};
	\draw (0,17.5) node (9) {};
	\draw (0,15.5) node (10) {};
	\draw (0,13.5) node (11) {};
	\draw (8,17.5) node (12) {};
	\draw (8,15.5) node (13) {};
	\draw (8,13.5) node (14) {};
	\draw (2,15) node (15) {};
	\draw (4,15.5) node (16) {};
	\draw (6,15) node (17) {};
	\draw (2,11.5) node (18) {};
	\draw (4,11) node (19) {};
	\draw (6,11.5) node (20) {};

	\draw (1)--(2)--(3)--(4)--(5);
	\draw (1)--(6)--(7)--(8)--(5);
	\draw (1)--(9)--(10)--(11);
	\draw (5)--(12)--(13)--(14);
	\draw (11)--(15)--(16)--(17)--(14);
	\draw (11)--(18)--(19)--(20)--(14);

	\draw (10,20) node (1b) {};
	\draw (12,22) node (2b) {};
	\draw (14,22.5) node (3b) {};
	\draw (16,22) node (4b) {};
	\draw (18,20) node (5b) {};
	\draw (12,18.5) node (6b) {};
	\draw (14,18) node (7b) {};
	\draw (16,18.5) node (8b) {};
	\draw (10,17.5) node (9b) {};
	\draw (10,15.5) node (10b) {};
	\draw (10,13.5) node (11b) {};
	\draw (18,17.5) node (12b) {};
	\draw (18,15.5) node (13b) {};
	\draw (18,13.5) node (14b) {};
	\draw (12,15) node (15b) {};
	\draw (14,15.5) node (16b) {};
	\draw (16,15) node (17b) {};
	\draw (12,11.5) node (18b) {};
	\draw (14,11) node (19b) {};
	\draw (16,11.5) node (20b) {};

	\draw (1b)--(2b)--(3b)--(4b)--(5b);
	\draw (1b)--(6b)--(7b)--(8b)--(5b);
	\draw (1b)--(9b)--(10b)--(11b);
	\draw (5b)--(12b)--(13b)--(14b);
	\draw (11b)--(15b)--(16b)--(17b)--(14b);
	\draw (11b)--(18b)--(19b)--(20b)--(14b);

	\begin{scope}[green]
	\draw (1b)--(3b) (1b)--(4b) (1b)--(5b) (1b)--(7b) (1b)--(8b);	
	\draw (2b)--(4b) (2b)--(5b) (2b)--(6b) (2b)--(7b) (2b)--(8b);
	\draw (3b)--(5b) (3b)--(6b) (3b)--(7b) (3b)--(8b);
	\draw (4b)--(6b) (4b)--(7b) (4b)--(8b);
	\draw (5b)--(6b) (5b)--(7b);
	\draw (6b)--(8b);
	
	\draw (11b)--(16b) (11b)--(17b) (11b)--(14b) (11b)--(19b) (11b)--(20b);	
	\draw (15b)--(17b) (15b)--(14b) (15b)--(18b) (15b)--(19b) (15b)--(20b);
	\draw (16b)--(14b) (16b)--(18b) (16b)--(19b) (16b)--(20b);
	\draw (17b)--(18b) (17b)--(19b) (17b)--(20b);
	\draw (14b)--(18b) (14b)--(19b);
	\draw (18b)--(20b);
	
	\end{scope}
	
	\draw (22,24) node (1c) {};
	\draw (24,26) node (2c) {};
	\draw (26,26.5) node (3c) {};
	\draw (28,26) node (4c) {};
	\draw (30,24) node (5c) {};
	\draw (24,22.5) node (6c) {};
	\draw (26,22) node (7c) {};
	\draw (28,22.5) node (8c) {};
	\draw (22,17.5) node (9c) {};
	\draw (22,15.5) node (10c) {};
	\draw (22,9.5) node (11c) {};
	\draw (30,17.5) node (12c) {};
	\draw (30,15.5) node (13c) {};
	\draw (30,9.5) node (14c) {};
	\draw (24,11) node (15c) {};
	\draw (26,11.5) node (16c) {};
	\draw (28,11) node (17c) {};
	\draw (24,7.5) node (18c) {};
	\draw (26,7) node (19c) {};
	\draw (28,7.5) node (20c) {};
	
	\draw (22,19.5) node (9d) {};
	\draw (22,13.5) node (10d) {};
	\draw (30,19.5) node (12d) {};
	\draw (30,13.5) node (13d) {};

	\draw (1c)--(2c)--(3c)--(4c)--(5c);
	\draw (1c)--(6c)--(7c)--(8c)--(5c);
	\draw (9d)--(9c)--(10c)--(10d);
	\draw (12d)--(12c)--(13c)--(13d);
	\draw (11c)--(15c)--(16c)--(17c)--(14c);
	\draw (11c)--(18c)--(19c)--(20c)--(14c);
	
	\begin{scope}[green]
	
	\draw (9d)--(12d) (10d)--(13d);

	\end{scope}

	\end{tikzpicture}
	
	\caption{A hole complex which after partial hole filling has a hole containing two paths.}
	
	\label{fig:overview}
\end{figure}

\end{document}